\documentclass[12pt,psfig]{article}
\usepackage{graphicx}
\begin{document}

\begin{center} 
{\bf  Self-consistent single-nucleon and single-$\Lambda$ potentials in strange nuclear matter        
with the Dirac-Brueckner-Hartree-Fock approach
 }              
\end{center}
\vspace{0.1cm} 
\begin{center} 
 F. Sammarruca \\ 
\vspace{0.2cm} 
 Physics Department, University of Idaho, Moscow, ID 83844, U.S.A   
\end{center} 
\begin{abstract}
We calculate self-consistent single-nucleon and single-$\Lambda$ 
potentials in hyperonic matter for different $\Lambda$ concentrations. 
The predictions include relativistic Dirac effects as reported previously
in a  calculation of the $\Lambda$ binding energy in symmetric nuclear matter. 
We discuss the dependence on momentum, density, and $\Lambda$ fraction. 
\\ \\ 
PACS number(s): 21.65.+f,21.80.+a 
\end{abstract}

\section{Introduction}

Studies of hyperon energies in nuclear matter are a useful starting point for 
calculations of bulk properties of 
hypernuclei. Furthermore, inclusion of strangeness in the equation of state is important towards a         
better understanding of the properties of matter inside neutron stars. 
This work is another step in our systematic exploring of nuclear and neutron matter 
under diverse conditions. 

Previously \cite{Sam08} we reported predictions of the $\Lambda$ binding energy 
in nuclear matter 
using the most recent meson-exchange nucleon-hyperon (NY) potential                 
from the  J{\"u}lich group \cite{NY05} (hereafter referred to as ``NY05") and including relativistic ``Dirac" effects 
on the $N\Lambda$ potential.            
In this paper, we will explore the behaviour of the self-consistently calculated single-$\Lambda$ and single-nucleon 
potentials with changing $\Lambda$ fraction. 

In our previous paper on the subject \cite{Sam08}, we discussed how 
some major quantitative differences between NY05 and the earlier version                
of the J{\"u}lich NY meson-exchange potential \cite{NY94}                           
impact predictions in dense matter. 
Accordingly, we expect large quantitative differences (beyond relativistic effects) when comparing with 
earlier microscopic calculations conducted                         
within the Brueckner-Hartree-Fock framework (see, for instance, 
Refs.~\cite{catania1,catania2}), which used        
the Nijmegen \cite{Nij89} and/or 
the J{\"u}lich \cite{NY89,NY94} NY meson-exchange potentials.

\section{The framework} 

For matter with non-vanishing hyperonic densities, 
the nucleon, $\Lambda$, and $\Sigma$ single-particle potentials are the solution of a    
 coupled self-consistency problem:                                    
\begin{eqnarray}
U_N = \int_{k<k_F^N} G_{NN} + \int_{k<k_F^{\Lambda}} G_{N \Lambda}                                       
 \int_{k<k_F^{\Sigma}} G_{N \Sigma}                                       
 \\ \nonumber                                              
U_{\Lambda} = \int_{k<k_F^N} G_{\Lambda N} + \int_{k<k_F^{\Lambda}} G_{\Lambda \Lambda}                             
+ \int_{k<k_F^{\Sigma}} G_{\Lambda \Sigma}                             
 \\ \nonumber                                              
U_{\Sigma} = \int_{k<k_F^N} G_{\Sigma N} + \int_{k<k_F^{\Lambda}} G_{\Sigma \Lambda}                             
+\int_{k<k_F^{\Sigma}} G_{\Sigma \Sigma}                             
\label{fs:eq9} 
\end{eqnarray}
In the above equations, $G_{NN}$, $G_{NY}$, and $G_{YY'}$,                                     
($Y,Y'=\Lambda,\Sigma$), are the nucleon-nucleon,      
nucleon-hyperon, and hyperon-hyperon $G$-matrices at some nucleon and hyperon densities 
defined by the Fermi momenta $k_F^N$ and $k_F^Y$. 

In the present calculation we consider a non-vanishing density of $\Lambda$'s but 
ignore the presence of real $\Sigma$'s in the medium 
(although both $\Lambda$ and $\Sigma$                        
are included in the coupled-channel calculation of the NY $G$-matrix, with             
free-space energies used for the latter). This scenario can be justified noticing that 
the (strong) reaction $N+\Sigma \rightarrow N+\Lambda$ 
is energetically always allowed, although it could be prevented by the Pauli principle at large 
hyperon concentrations. 
Thus, over a time scale which is long relative to 
strong processes, and for small hyperon concentrations,                                              
equilibrated matter can be reasonably assumed to contain mostly nucleons and $\Lambda$'s
\cite{catania1}. Also, we neglect the $YY'$ interaction, as very little information
is available about it. For these reasons, we keep the $\Lambda$ concentration           
relatively low. 
                       
We start with symmetric nuclear matter at some 
Fermi momentum $k_F^N$ in the presence of a ``$\Lambda$ impurity", i.e.                  
$k_F^{\Lambda}\approx 0$ and then increase the $\Lambda$ concentration.  
The parameters of both the nucleon and the $\Lambda$ potentials are calculated self-consistently
with the $G_{NY}$ and the $G_{NN}$ interactions, which are the solution of the Bethe-Goldstone 
equation with $NY$ and $NN$ potentials, respectively. We solve the coupled self-consistency 
problem above using the same techniques as described  previously for isospin-asymmetric matter \cite{AS03}.
In the Brueckner calculation, 
density-dependent effects come in through 
angle-averaged Pauli blocking and dispersion. Most of the details for the nucleon-nucleon sector are given in Ref.~\cite{AS03}.             
For two particles with masses $M_N$ and $M_Y$, ($Y=\Lambda,\Sigma$), and Fermi momenta
$k_F^N$ and $k_F^Y$, Pauli blocking requires               
\begin{equation}
\frac{(\frac{M_N}{M}P)^2 +k^2-(k_F^N)^2}{2Pk\frac{M_N}{M}} > cos \theta >
-\frac{(\frac{M_Y}{M}P)^2 +k^2-(k_F^Y)^2}{2Pk\frac{M_Y}{M}}.                   
\end{equation} 
where $\theta$ is the angle between the total (${\vec P}$) and the relative
(${\vec k}$) momenta of the two particles, and $M=M_Y + M_N$. Angle-averaging is then applied in the 
usual way. 

Dirac effects are applied in the $NN$ as well as the 
$N\Lambda$ potentials. 
This entails involving the 
$\Lambda$ single-particle Dirac wave function in the self-consistent calculation through the 
$\Lambda$ effective mass \cite{Sam08}.                
For the nucleon-nucleon sector, we use the Bonn B potential \cite{Mac89}, which is a relativistic one-boson exchange 
potential developed within the context of the Thompson equation and uses the pseudovector
coupling for pseudoscalar mesons. 
The problem associated with the DBHF approach and the use of the pseudoscalar coupling 
for the interactions of 
pions and kaons with nucleons and hyperons, (which occurs for the                                        
J{\"u}lich NY potential), was discussed                                                             
in Ref.~\cite{Sam08}.

\begin{figure}
\begin{center}
\vspace*{-4.0cm}
\hspace*{-2.0cm}
\scalebox{0.7}{\includegraphics{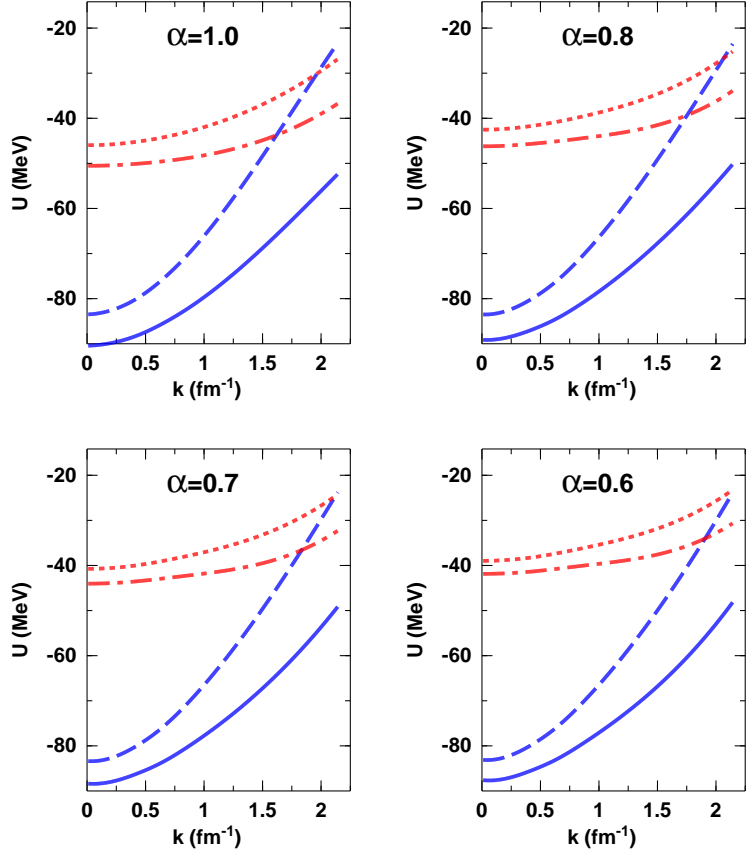}}
\vspace*{-3.0cm}
\caption{Single-nucleon and single-$\Lambda$ potentials as a function of the momentum
at constant total density and for varying $\Lambda$ concentrations. The dotted and the 
dashed lines are DBHF predictions for the $\Lambda$ and the nucleon potentials, respectively,
while the dash-dotted and the solid curves are the corresponding results without the Dirac effects. 
See text for more details. 
} 
\label{one}
\end{center}
\end{figure}

\section{The single-particle potentials}                                

We show in Fig.~1 the single-nucleon and the single-$\Lambda$ potentials as a function of 
the momentum. The first panel corresponds to a nucleon Fermi momentum of 1.35$fm^{-1}$ and zero
density of $\Lambda$ particles, that is, the situation of a $\Lambda$ impurity in nuclear matter.
Notice that this implies 
\begin{equation}
k_F^N = k_F(1 + \alpha)^{1/3}=1.35fm^{-1}, 
\end{equation}
and 
\begin{equation}
k_F^{\Lambda} = k_F(1 - \alpha)^{1/3}=0, 
\end{equation}
where $k_F$ is the average Fermi momentum  
and $\alpha = \frac{\rho_N-\rho_{\Lambda}}{\rho_N+\rho_{\Lambda}}=1$.                                       
The other three cases correspond to $\alpha$=0.8, 0.7, and 0.6,          
keeping the average Fermi momentum 
constant. In terms of $\Lambda$ fraction, $Y_{\Lambda}$, the above $\alpha$ values correspond to 
$Y_{\Lambda}$=$0$, $0.1$, $0.15,$, and $0.2$, respectively.                                 
In all panels, the two lower curves (solid and dash) are the results of a conventional Brueckner calculation (solid)        
and of a DBHF 
calculation (dash) for the nucleon potential, whereas the higher curves have the same meaning for 
the $\Lambda$ single-particle potentials, with the dotted lines being the Dirac predictions. 
In the first panel of 
Fig.1, the value of $U_{\Lambda}$ at $k$=0 is of course the $\Lambda$ binding energy we
calculated in Ref.~\cite{Sam08}. The Dirac effect is small at low momentum but increases
considerably with increasing momentum, and even more so for the nucleon. 
\begin{figure}
\begin{center}
\vspace*{-4.0cm}
\hspace*{-2.0cm}
\scalebox{0.4}{\includegraphics{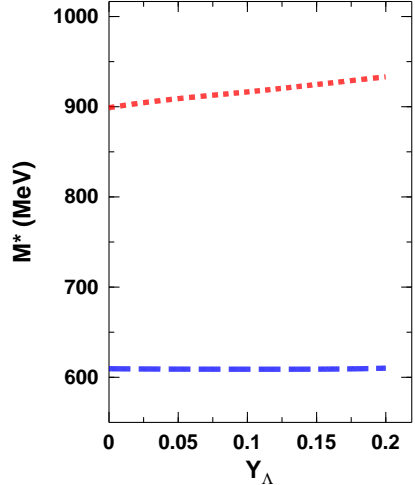}}
\vspace*{-1.5cm}
\caption{The nucleon (dashed line) and the $\Lambda$ (dotted) effective masses versus the                         
$\Lambda$ fraction at fixed density. 
The predictions correspond to the dashed and dotted lines in Fig.~1.                             
} 
\label{two}
\end{center}
\end{figure}

\begin{figure}
\begin{center}
\vspace*{-4.0cm}
\hspace*{-2.0cm}
\scalebox{0.4}{\includegraphics{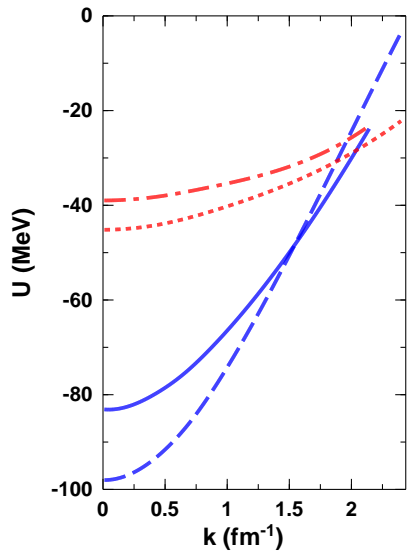}}
\vspace*{-1.5cm}
\caption{DBHF predictions for the single-nucleon and the single-$\Lambda$ potentials as a function of the momentum
at two different densities and fixed $\Lambda$ concentrations. The dotted(dash-dotted) and the 
dashed(solid) lines refer to the $\Lambda$ and the nucleon potentials, respectively, at the larger(smaller) density. 
See text for more details. 
} 
\label{three}
\end{center}
\end{figure}

With increasing $\Lambda$ concentration, the total density remaining constant, both potentials 
become less attractive, the effect being more pronounced for the $\Lambda$.                              
Consistent with Fig.~1, the $\Lambda$ 
effective mass becomes larger with increasing $Y_{\Lambda}$ (although very slowly), while the nucleon 
effective mass is considerably less sensitive to these (small) variations, see Fig.~2. 

The results shown in Fig.~1 can be understood 
observing that, as 
the $\Lambda$ fraction increases, $U_{\Lambda}$ will receive contribution from a smaller number of nucleons
(the $\Lambda \Lambda$ interaction is neglected, see second line of Eq.~(1)), whereas $U_N$ will have a slightly reduced 
contribution from $G_{NN}$ and a slightly enhanced contribution from $G_{N \Lambda}$, see first 
line of Eq.~(1). 
Also, the Dirac effect becomes smaller with increasing $\Lambda$ concentration, 
which is also plausible since this effect is larger for nucleons. 

Comparing with the microscopic predictions from Ref.~\cite{catania1}, we see that 
the qualitative features are similar, but there are very large quantitative differences
due to the use of different $NY$ potentials, the depth of 
$U_{\Lambda}$ at zero $\Lambda$ density being about -30 MeV in Ref.~\cite{catania1}. 

In Fig.~3 we compare the single-nucleon and the single-$\Lambda$ potentials (DBHF predictions
only) at two different densities, corresponding to a nucleon                                    
Fermi momentum equal to 1.25$fm^{-1}$ and 1.4$fm^{-1}$, respectively, with 
the $\Lambda$ fraction kept fixed at $20\%$.                 
Both potentials start deeper at the larger density.               
As already clear from Fig.~1,
$U_N$ grows repulsive (as a function of the momentum) at a much faster rate than $U_{\Lambda}$. Figure 3 
shows that this is more so the larger the density, 
most likely due to the (very density-dependent) Dirac
effect, which impacts nucleons more strongly.

\section{Conclusions }                                
We have reported  non-relativistic and                               
Dirac-Brueckner-Hartree-Fock predictions of the 
$\Lambda$ and nucleon single-particle potentials in nuclear matter at moderately high densities.  
We examined the significance of Dirac effects as a function of momentum and $\Lambda$
concentration. 
Dirac effects on hyperons may be important at the higher densities, where high momenta are probed. 

We have neglected the $\Lambda \Lambda$ interaction, which is not a serious problem at low
$\Lambda$ concentrations. The hyperon fraction in $\beta$-equilibrated matter is of course                           
density dependent, but we expect that it will 
remain relatively small, especially if the 
$NY$ interaction is fully
taken into account (as compared to a model of non-interacting hyperons).     
This is because the (repulsive) short-range $NY$ interaction will become effective at high density 
and, to a certain degree, may prevent 
the growth of the hyperon fraction (at the expenses of the lepton population) which has been 
observed in models of non-interacting hyperons \cite{BBS00}.

Hypernuclei can be produced in large number using relativistic heavy ion collisions \cite{BP07}.
Calculations such as the present one can be useful as a foundation for studies          
of hypernuclei based on generalized mass formulas \cite{SCB06} and thus                           
predictions of stability of systems where one or more nucleons
are replaced with $\Lambda$'s. Since the energy of a single $\Lambda$ impurity in 
symmetric matter (at normal nuclear matter density) is about -50 MeV (or about -30 MeV, 
depending on the $NY$ potential being used), whereas the average energy of a nucleon in 
nuclear matter is approximately -16 MeV, one may conclude that it is energetically favorable to replace a 
single nucleon with a single $\Lambda$. Of course this may no longer be true with increasing 
$\Lambda$ fraction, as the $\Lambda$ particles will have to acquire kinetic energy to obey the Pauli 
principle. This and other issues will be explored further in        
forthcoming work.

\section*{Acknowledgments}
Support from the U.S. Department of Energy under Grant No. DE-FG02-03ER41270 is 
acknowledged.                                                                           


\begin{thebibliography}{99}
\bibitem{Sam08} F. Sammarruca, arXiv:0801.0879 [nucl-th]. 
\bibitem{NY05} J. Haidenbauer and Ulf-G. Meissner, Phys. Rev. C {\bf 72}, 044005 (2005).
\bibitem{catania1} H.-J. Schulze, M. Baldo, and U. Lombardo, Phys. Rev. C {\bf 57}, 704 (1998).
\bibitem{catania2} M. Baldo, G.F. Burgio, and H.-J. Schulze, Phys. Rev. C {\bf 58}, 3688 (1998).
\bibitem{Nij89} P. Maessen, Th. Rijken, and J. de Swart, Phys. Rev. C {\bf 40}, 2226 (1989). 
\bibitem{NY89} B. Holzenkamp, K. Holinde, and J. Speth, Nucl. Phys. {\bf A500}, 485    
              (1989). 
\bibitem{NY94} A. Reuber, K. Holinde, and J. Speth, Nucl. Phys. {\bf A570}, 543 (1994). 
\bibitem{AS03} D. Alonso and F. Sammarruca, Phys. Rev. C {\bf 67}, 054301 (2003). 
\bibitem{Mac89} R. Machleidt, Adv. Nucl. Phys. {\bf 19}, 189 (1989). 
\bibitem{BBS00} M. Baldo, G.F. Burgio, and H.-J. Schulze, Phys. Rev. C {\bf 61}, 055801 (2000). 
\bibitem{BP07} A.S. Botvina and J. Pochodzalla, Phys. Rev. C {\bf 76}, 024909 (2007).
\bibitem{SCB06} C. Samanta, P. Roy Chowdhury, and D.N. Basu, J. Phys. G {\bf 32}, 363 (2006). 
\end{thebibliography}
\end{document}